\documentclass[11pt]{article}
\usepackage{marginnote}

\usepackage{color}
\usepackage{amsmath}
\usepackage{amsthm}
\numberwithin{equation}{section}

\usepackage{array}
\usepackage{tabularx}
\usepackage{appendix}
\usepackage{makecell}
\usepackage{graphicx}
\usepackage{amssymb}
\usepackage[T1]{fontenc}
\usepackage[utf8]{inputenc}
\usepackage[raggedright]{titlesec}
\usepackage{blindtext}
\usepackage{commath}
\usepackage{cite}
\usepackage{caption}
\usepackage[textwidth = 7in]{geometry}
\usepackage{mathtools}
\usepackage[dvipsnames]{xcolor}
\usepackage{enumitem}
\usepackage{amsmath}
\usepackage{geometry}
\usepackage{lipsum}  
\usepackage{thmtools}
\usepackage{setspace}
\titleformat{\paragraph}[hang]{\normalfont\normalsize\bfseries}{\theparagraph}{1em}{}
\titlespacing*{\paragraph}{0pt}{3.25ex plus 1ex minus .2ex}{0.5em}
\makeatletter
\DeclareRobustCommand{\change}{%
	\@bsphack
	\leavevmode
	\color{magenta}%
	\@esphack
}
\DeclareRobustCommand{\stopchange}{%
	\@bsphack
	\normalcolor
	\@esphack
}
\makeatother

\stepcounter{secnumdepth}
\stepcounter{tocdepth}

\usepackage{hyperref}
\hypersetup{
  colorlinks   = true, 	
  urlcolor     = blue, 	
  linkcolor    = blue, 	
  citecolor   = magenta 	
}

\title{
	\vskip-1.3cm
Dirac Equation with Space Contributions Embedded in a Quantum-Corrected Gravitational Field}

\author{
	M. Baradaran$^1$\footnote{marzieh.baradaran@uhk.cz, ORCID: \href{http://orcid.org/0000-0002-8455-9973}{0000-0002-8455-9973}}, 
	L.M. Nieto$^2$\footnote{luismiguel.nieto.calzada@uva.es, ORCID: \href{http://orcid.org/0000-0002-2849-2647}{0000-0002-2849-2647}}, 
	and 
	S. Zarrinkamar$^{2,3}$\footnote {saber.zarrinkamar@uva.es, ORCID: \href{http://orcid.org/0000-0001-9128-4624}{0000-0001-9128-4624}}
	\\  [1ex]
	\small
	$^1$\,Department of Physics, Faculty of Science, University of Hradec Kr\'alov\'e,\\
	\small
	Rokitansk\'eho 62, 500 03 Hradec Kr\'alov\'e, Czechia
	\\ [1ex]
	\small
	$^2$\,Departamento de F\'{\i}sica Te\'{o}rica, At\'{o}mica y \'{O}ptica, and Laboratory for Disruptive \\ 
		\small
		Interdisciplinary Science (LaDIS), Universidad de Valladolid, 47011 Valladolid, Spain
	\\ [1ex]
	\small
	$^3$\,Departament of Basic Sciences, Garmsar Branch,
	Islamic Azad University, Garmsar, Iran
}

\begin{document}
	
	\maketitle

\begin{abstract}

The Dirac equation is considered with the recently proposed generalized gravitational interaction (Kepler or Coulomb), which includes post-Newtonian (relativistic) and quantum corrections to the classical potential. The general idea in choosing the metric is that the spacetime contributions are contained in an external potential or in an electromagnetic potential which can be considered as a good basis for future studies of quantum physics in space. The forms considered for the scalar potential and the so-called vector (magnetic) potential, can be viewed as the multipole expansion of these terms and therefore the approach includes a simultaneous study of multipole expansions to both fields. We also discuss several known generalizations of the Coulomb potential within this formulation in terms of certain Heun functions. The impossibility of solving our equation for the quantum-corrected Coulomb terms using known exact or quasi-exact nonperturbative analytical techniques is discussed, and finally the Bethe-ansatz approach is proposed to overcome this challenging problem.

\end{abstract}


\section{Introduction}
The study of quantum mechanics in non-Minkowskian spacetimes has been pursued for decades with various motivations, including the desire to find a relationship between quantum theory and general gravity \cite{Hollands, Belenchia 2022}.
Interest in this field has been revived by the remarkable applications of the curved spacetime formulation in particle and solid state physics \cite{Vozmediano}, as well as by the opening of recent lines of research in both quantum optics and quantum technologies \cite{Genov 2009, Schultheiss 2010, Bruschi 2014, Bekenstein 2015, Schultheiss 2016}.
This is, of course, part of the beauty of physics, where seemingly uncorrelated fields can be inherently related. The reviews entitled ``Quantum fields in curved space-time'' \cite{Hollands}, ``Quantum physics in space'' \cite{Belenchia 2022} and ``Gauge fields in graphene'' \cite{Vozmediano} provide an excellent exposition of the physical and mathematical foundations needed to address this topic. In the following, we will review some more recent research in this field, which will allow us to properly contextualize our future work.

Starting with quantum condensed matter applications, let us note that solutions of the Dirac equation were derived using the so-called frozen Gaussian approximation and the results were applied to investigate scattering on deformed graphene surfaces \cite{Chai}.
In pioneering work, Genov and his collaborators linked strange artificial materials, called metamaterials, which find exceptional applications in modern quantum technologies, with the curved formalism of spacetime, a task that might seem strange at first glance \cite{Genov 2009}.
Motivated by the recent revisit of fundamental physics ideas in relation to quantum technologies, the current state of space quantum technologies is reviewed in \cite{Belenchia 2022, Mohageg 2022}. The structure of slightly curved graphene sheets was analyzed by Cortijo and Vozmediano \cite{Cortijo} and Neto et al. \cite{Castro} and the low-energy properties of graphene wormholes were derived using the curved space Dirac equation in \cite{Gonzalez}. A comprehensive review of graphene gauge fields in curved spacetime was performed in \cite{Vozmediano}.
The continuum limit of time-dependent and space-dependent discrete-time quantum walks was studied based on an analogy with the propagation of a massless Dirac particle in a gravitational field \cite{Molfetta}.
A cylindrical graphene sample was analytically studied within the framework of a curved spacetime Dirac equation using an appropriate Kubo formula \cite{Gallerati}. The general belief regarding the curved space description of elastically deformed monolayer graphene was seriously criticized in \cite{Roberts}, where the authors claim that such model is valid only for special and extremely tight conditions.
Various aspects of the Dirac equation for curved spacetime were discussed at the graduate level in Collas's instructive book \cite{Collas}.
The change in atomic levels imposed by high intensity lasers and the connection with the Dirac equation of curved space was studied in \cite{Toma}. Bagchi and his collaborators studied the motion of charge carriers in curved Dirac materials when a local Fermi velocity \cite{Bagchi} is present.

As regards the domain of optics and quantum optics, we can mention that Schultheiss experimentally analyzed the effect of the curvature of space on the evolution of light for surfaces with constant Gaussian curvature \cite{Schultheiss 2010} and
Bruschi et al. investigated the consequences of curved spacetime on space-based quantum communication protocols \cite{Bruschi 2014}.
Gravitational effects were simulated in \cite{Bekenstein 2015} with optical wave packets and long-range nonlinearities, which mathematically resemble the Newton-Schr\"odinger system.
Again, Schultheiss and others reviewed Hanbury Brown and Twiss measurements in curved space and commented on the analogy with cosmological models \cite{Schultheiss 2016}. Using a dual light-pulse atomic interferometer, Asenbaum et al. reported the phase shift due to tidal force \cite{Asenbaum 2017}. 
Starting from basic concepts of general relativity, some nanophotonic structures were proposed in which the temporal evolution is controlled by the spatial curvature of the medium \cite{Bekenstein 2017}. 
Using the Huygens-Fresnel principle, Xu and Wan formulate a basis for studying light propagation on 2D surfaces of arbitrary curvature \cite{Xu 2021} and using the wave equation corresponding to curved space, a generalized formula for the light path in curved space is obtained in \cite{Shao 2021}.
Using transformation optics, Sheng et al. studied various transformation structures in a curved spatial approach \cite{Sheng 2023}. The phenomenon of branched light flux on surfaces of Gaussian curvature was studied in \cite{Ding 2023} and the effect of a weak gravitational field on optical solitons in a nonlinear medium was analyzed in \cite{Spengler 2023}. 
The confinement problem of Dirac fermions in a two-dimensional curved space was investigated in \cite{Flouris}, while in \cite{Pedernales} an exact mapping between the Dirac equation of (1+1)-dimensional curved spacetime and a multiphoton Rabi model is addressed.

From the point of view of mathematical physics, there are interesting papers investigating various wave equations, including the Dirac equation in curved space.
The anomalous properties of Dirac particles in curved space were studied by Griffith \cite{Griffiths} and the Dirac theory in a curved Riemann-Cartan frame was formulated by Nieh and Yan \cite{Nieh}.
The Dirac equation in one spatial dimension was solved for the Robertson-Walker spacetime and the Cigar metrics in \cite{Sinha} and the two-dimensional Dirac equation in the presence of an electromagnetic field was solved by Panahi and Jahangiri \cite{Panahi}, while the motion of rotating particles in curved spacetime was analyzed in \cite{dAmbrosi2015}.
Maxwell's equations were reformulated in anti-de Sitter spacetime in \cite{Herdeiro 2015}. In \cite{Hollands} a comprehensive review of various quantum fields in curved spacetimes was carried out and in \cite{Noble} Dirac fermions in curved spacetime originating from a massive charged object were studied for the Reissner-Nordstr\"om case.
Kobakhidze and his collaborators analyzed the Zitterbewegung effect in a curved spacetime frame \cite{Kobakhidze 2016}. Winstanley \cite{Winstanley} investigated the stability of sphaleronic black holes in the Einstein-Yang-Mills-Higgs theory with a Higgs doublet in anti-de Sitter spacetime.
As far as quantum models are concerned, the one-dimensional Dirac equation in curved spacetime was solved analytically by Ghosh and Roy and the existence of bound states was discussed for various cases \cite{Ghosh2021}, and  the hydrogen atom, Morse oscillator and a linear radial potential were solved within the framework of curved spacetime Dirac equation by Oliveira and Schmidt \cite{Oliviera}. The Dirac equation modified with constant curvature was solved for the Dunkl oscillator in \cite{Sedaghatnia} and the Dirac equation with a position-dependent mass was investigated in a two-dimensional curved-space for PT/non-PT-symmetric potentials including Morse and Scarf potentials in  \cite{Yesiltas}. 
The correspondence between the one dimensional curved spacetime Dirac equation and a nonlinear Schrödinger model was commented on by a $(1+1)-$dimensional bosonization in \cite{Ghosh}. Holography in anti-de Sitter space was studied using an integral transformation approach \cite{Bhowmick 2019}. The analogy between the $(1+1)$ and $(2+1)$ curved spacetime Dirac equation with spacetime-dependent quantum walks was addressed in \cite{Mallick, Arighi}.
The Dirac oscillator problem in curved spacetime was revisited in the so-called spin and pseudospin symmetry limits \cite{Oliviera2019}. Solutions of the Dirac equation in curved space were reported in a novel approach by Yagdijan and Galstian in connection with the generalized Euler-Poisson-Darboux equation via an integral transformation approach \cite{Yadijan}.
Using a Madelung transformation, the hydrodynamic representation of the Dirac equation in curved spacetime was formulated \cite{Matos 2022}.
The problem of locality of higher spin gravity in Euclidean anti-de Sitter space was exhaustively analyzed by Neiman \cite{Neiman}.
Using the Wentzel-Kramers-Brillouin approximation, the Dirac equation was analyzed in a curved background and several concepts  were studied, including the spin Hall effect and the Berry curvature \cite{Oaneca}.
On the other hand, very recently, a generalized gravitational potential is proposed which includes an inverse square term as the relativistic correction as well as an inverse cubic term as the dominant quantum correction to the classical term \cite{Netto, Fr, Lambiase}. 
This can be thought of, in a conceptual sense, as a kind of unified interaction, broadly speaking, and therefore it will be interesting to consider the new potential within a special form of curved space.

The structure of the present study is as follows. In Section~\ref{sec2} we first review the most essential formulae of the Dirac equation in the desired metric.  Using the Bethe-ansatz approach, in Section~\ref{sec:genSol} we report the general solution for arbitrary $n$, while solutions for the ground state and the first excited state are presented in Sections~\ref{sect:Grn0} and \ref{sect:Grn01}, respectively, being the special case of the Coulomb problem also derived in Section~\ref{sect:CoulombGrn0}. We have included Appendix~\ref{appendix} to discuss the impossibility of solving our problem using other commonly used nonperturbative analytical techniques, and thus clarify why we use the Bethe-ansatz approach \cite {Zhang AoP 2012, Zhang 2012, Baradaran, Zhang 2024 AOP}.
The article ends with some final conclusions.

\section{Curved spacetime Dirac equation with a generalized metric}\label{sec2}

Let us start by remembering the fundamental formulas that will be necessary, in order to preserve the continuity of our work with some of the references already mentioned. The Dirac equation \cite{Collas}
\begin{equation} 
\left(i\gamma^\mu\nabla_\mu -mc\right)\Psi=0,
\end{equation}
will be considered for the metric
\begin{equation}
g_{\mu\nu}=\text{diag}\left(e^{2f(r)},-e^{2g(r)},-r^2,-r^2\sin^2\theta\right),
\end{equation}
where, in principle, $f(r)$ and $g(r)$ are arbitrary functions of the radial coordinate, and it is clearly seen that the angular parts are the same as in Minkowski spacetime.
Recalling that in curved spacetime $\nabla_\mu=\partial _{\mu}+{iA_\mu}/c+\Omega_\mu$, where $\partial _{\mu}$ is the covariant derivative on flat spacetime and $\Omega_\mu$ denotes the spin connection, and choosing $A_\mu=(V(r),cA_r(r),0,0)$, after some calculation, which are developed completely  in \cite{Oliviera}, we can write the spinor wave function as
\begin{equation}
\Psi_c(r,\theta,\phi)=N\left( \begin{array}{c} R_1(r)\ \mathcal{Y}_{j+1/2}^{|m|\, j}(\theta,\phi) \\ [1.2ex]
 iR_2(r)\ \mathcal{Y}_{j-1/2}^{|m|\, j}(\theta,\phi)\end{array} \right),
\end{equation}
where $N$ is the normalization constant and 
\begin{equation}
\mathcal{Y}_{l}^{j=l\pm1/2\, m}(\theta,\phi) =\frac{1}{\sqrt{2l+1}}\left( \begin{array}{c}{\pm\sqrt{l\pm m+\frac{1}{2}}} \ Y_{l}^{m-1/2}(\theta,\phi) \\  [1.2ex]
{\sqrt{l\mp m+\frac{1}{2}}}\ Y_{l}^{m+1/2}(\theta,\phi)\end{array} \right)
\end{equation}
are the spinor spherical harmonics, being $Y_{l}^{m}(\theta,\phi)$ the usual spherical harmonics.
From now on we will take $m=\hbar=1$, $c=1/\alpha$, with $\alpha$ being the fine structure constant. 
Considering $f(r)=g(r)$, with $e^{f(r)}=1+\alpha^2U(r)$, and introducing $R_1(r)=\frac{u(r)}{r}e^{- {f(r)}/{2}}$ and $R_2(r)=\frac{v(r)}{r}e^{- {f(r)}/{2}}$, we obtain
\begin{equation}\label{xyz}
\left(\begin{array}{cc}{1+\alpha^2\left(V(r)+U(r)\right)}&{-\alpha \left[\frac{d}{dr}-\frac{\lambda}{r}\left(1+\alpha^2 U(r)\right)-A_r (r)\right]} \\ [1ex]
{\alpha \left[\frac{d}{dr}+\frac{\lambda}{r}\left(1+\alpha^2 U(r)\right)+A_r (r) \right]}&{-1+\alpha^2\left(V(r)-U(r)\right)}\end{array}\right) \left( \begin{array}{c} u(r)\\  [1ex]
v(r)\end{array} \right)=\epsilon\left( \begin{array}{c} u(r)\\  [1ex]
v(r)\end{array} \right).
\end{equation}
By introducing the transformation $U'=\text{exp}(i\eta\sigma_2)$, where $-\tfrac{\pi}2\leq\eta\leq\tfrac{\pi}2$ and $\sigma_2$ denotes the second Pauli matrix, the equation \eqref{xyz} appears as
\begin{equation}
\left(\begin{array}{cc}{C+2\alpha^2V(r)-\epsilon}&{-S+\frac{\alpha^2}{S}\left(CV(r)-U(r)\right)-\alpha \frac{d}{dr}}\\  [1ex]
-S+\frac{\alpha^2}{S}\left(CV(r)-U(r)\right)+\alpha \frac{d}{dr} & -C-\epsilon
\end{array}\right) \left( \begin{array}{c} \rho_1(r) \\  [1ex]
\rho_2(r)\end{array} \right)=0,
\end{equation}
where  
\begin{equation}
\left( \begin{array}{c} \rho_1(r)  \\ \rho_2(r)\end{array} \right)=U'\left( \begin{array}{c} u(r)  \\  v(r)\end{array} \right), 
\quad 
C=\cos2\eta 
\quad \text{and} \quad 
S=\sin2\eta. 
\end{equation}
Eliminating one component in favor of the other using $A_r(r)=\frac {\alpha C}{S}\left[\frac{V(r)}{C}-U(r)\right]-\frac{\lambda}{r}\left[1+\alpha^2 U(r)\right]$, we find that the final equation is \cite{Oliviera}
\begin{equation}\label{rho1}
\left[\frac{d^2}{dr^2}+\frac{\alpha}{S} \frac{d\left(CV(r)-U(r)\right)}{dr}-2\left(U(r)+\epsilon V(r)\right)-\frac{\alpha^2}{S^2}\left(CV(r)-U(r)\right)^2+\frac{\epsilon^2-1}{\alpha^2}\right]\rho_1(r)=0,
\end{equation}
while the other component is simply obtained by 
\begin{equation}
\rho_2(r)=\frac{\alpha}{C+\epsilon}\left[-\frac{S}{\alpha}+\frac{\alpha}{S}\left(CV(r)-U(r)\right)+\frac{d}{dr}\right]\rho_1(r).
\end{equation}
Simplifying equation \eqref{rho1} by $V(r)=az(r)$ and $U(r)=bz(r)$, without loss of generality in our case, we arrive at \cite{Oliviera}
\begin{equation}\label{CurDirEq}
\left[\frac{d^2}{dr^2}+\frac{\alpha}{S}(aC-b)z'(r)-2(b+\epsilon a)z(r)-\frac{\alpha^2}{S^2}(aC-b)^2z^2(r)+\frac{\epsilon^2-1}{\alpha^2}\right]\rho_1(r)=0.
\end{equation}
Obviously, using the above relations, the original components can be derived in a straightforward manner. 

For completeness, at the end of the paper we have included a short Appendix~\ref{appendix} explaining why we use the Bethe-ansatz approach below, and where we attempt to provide a conceptual overview of frequently used analytical techniques \cite{El-Jaick, Ronveaux, Ishkhanyan 2018, Turbiner, Turbiner88, Artemio94, Arfken}, all of which fail in our case, to the best of our knowledge.

\section{Solutions by the Bethe-ansatz approach}
\label{sect:BAEZhang}

Based on the previous sections, we now know that we need to consider another approach to better understand the quasi/conditionally exact analytical solutions of our problem. Let us consider $z(r)$ in \eqref{CurDirEq} as follows
\begin{equation}\label{zr}
	z(r)= \frac{u}{r}+\frac{v}{r^2}+\frac{w}{r^3}, \qquad u,v,w<0.
\end{equation}
It is worth mentioning that, to our knowledge, this potential has only been investigated within the framework of the generalized uncertainty principle \cite{Baradaran}. Substituting into \eqref{CurDirEq}, we arrive at a Schr\"odinger-like equation
\begin{equation}\label{eqDirac}
	\left\{ \frac{d^2}{dr^2}+ \frac{\epsilon_n ^2-1}{\alpha ^2}-\frac{2 u (a \epsilon_n +b)}{r}+\frac1{S^2}\left(\frac{\Lambda_2}{r^2}+\frac{\Lambda_3}{r^3}+\frac{\Lambda_4}{r^4}\right)-\frac{\alpha ^2 (b-a C)^2}{S^2}\left(\frac{w^2}{r^6}+\frac{2 v w}{r^5}\right)\right\}\rho_{1,n}(r)=0,
\end{equation}
where
\begin{equation}\label{lambdas}
	\begin{aligned}
		\Lambda_2=&   -b \left(2 S^2 v-2 a \alpha ^2 C u^2-\alpha  S u\right)-a \alpha  C u (a \alpha  C u+S)-2 a S^2 v \epsilon_n -\alpha ^2 b^2 u^2            \, ,\\[1pt]
		\Lambda_3=&  -2 \left(b \left(-2 a \alpha ^2 C u v+S^2 w-\alpha  S v\right)+a \alpha  C v (a \alpha  C u+S)+a S^2 w \epsilon_n +\alpha ^2 b^2 u v\right)   \,,\\[1pt]
		\Lambda_4= 	& \alpha  (b-a C) \left(3 S w-\alpha  (b-a C) \left(2 u w+v^2\right)\right)    \, ,
	\end{aligned}
\end{equation}
where we have added the index $n$ to $\epsilon$ and $\rho_{1}(r)$ to distinguish some states from others. Now, we propose
\begin{equation}\label{WavAns}
	\rho_{1,n}(r)=e^{\Delta(r)}\,\mathcal{R}_{1,n}(r) ,\qquad \Delta(r)=\delta \ln r+\frac{\gamma }{r^2}+\frac{\beta }{r}+\lambda  r ,
\end{equation}
where the new function $\mathcal{R}_{1,n}(r)$ is a polynomial, and $\lambda\,,\beta\,,\gamma<0$ and $\delta>0$ are parameters to be determined. Substituting \eqref{WavAns} into \eqref{eqDirac} and solving the resulting Riccati equation for $\Delta(r)$, we obtain
\begin{subequations}\label{parameters}
	\begin{align}
		\delta&=  \frac32+\frac{\alpha  w (b-a C) \left(2 \alpha  u (b-a C)-3 S\right)}{2\sigma  S^2}   ,\\
		\lambda&=  -\sqrt{\frac{1-\epsilon_n ^2}{\alpha ^2}}\, ,\quad
		\gamma=  -\frac{\sigma }{2}  ,\quad
		\beta=- \frac{ v}{w}\,\sigma \,,
	\end{align}
\end{subequations}
in which
\begin{equation}\label{sigma}
	\sigma=\sqrt{\frac{\alpha ^2 w^2 (b-a C)^2}{S^2}} .
\end{equation}
Consequently, the differential equation for $\mathcal{R}_{1,n}(r)$ simplifies to
\begin{equation}\label{eq332}
	\left\{r^3\frac{d^2}{dr^2}+\left( 2 \lambda  r^3+2 \delta  r^2-2 \beta  r-4 \gamma\right)\frac{d}{dr}+\left(\xi_2 \,r^2+\frac{\xi_1}{S^2} \,r+\frac{\xi_0}{S^2}\right)\right\}\mathcal{R}_{1,n}(r)=0,
\end{equation}
where
\begin{equation}
	\begin{aligned}
		\xi_0=& -2 a^2 \alpha ^2 C^2 u v-2b \left(S^2 w-2 a \alpha ^2 C u v-\alpha  S v\right)-2a S \left(\alpha  C v+S w \epsilon_n \right)-2\alpha ^2 b^2 u v           \\
		& -2S^2 \left(\beta  (\delta -1)+2 \gamma  \lambda \right)           \, ,\\[6pt]
		\xi_1=&  -a^2 \alpha ^2 C^2 u^2+b \left(2 a \alpha ^2 C u^2-2 S^2 v+\alpha  S u\right)-a S \left(\alpha  C u+2 S v \epsilon_n \right)-\alpha ^2 b^2 u^2        \, \\
		& +S^2 \left((\delta -1) \delta -2 \beta  \lambda \right)          \, ,\\[6pt]
		\xi_2= &   2 \delta  \lambda -2 u \left(a \epsilon_n +b\right)   .
	\end{aligned}
\end{equation}
To solve \eqref{eq332}, we now assume polynomial solutions for $\mathcal{R}_{1,n}(r)$ of the form \cite{Zhang}
\begin{equation}\label{ansatZhang}
	\mathcal{R}_{1,n}(r)=\! \left\{
	\begin{array}{cl}
		1, & \quad n=0, \\ 
		\displaystyle\prod_{i=1}^n (r-r_i), & \quad n\in\mathbb{N},
	\end{array}
	\right.  
\end{equation}
where $r_i$ are distinct roots to be determined. The general solutions of \eqref{eq332} are given by the following set of equations  
\begin{subequations}\label{BAEgeneqs}
	\begin{align}
		& 	\xi_2+2n\lambda =0 ,\label{BAEgeneqs1}\\
		& 	\xi_1+2\lambda \sum_{i=1}^{n} r_i+n(n-1) +2n\delta=0 ,\label{BAEgeneqs2}\\
		& \xi_0+2\lambda \sum_{i=1}^{n} r_i^2+ 2 (\delta+n-1)\sum_{i=1}^{n} r_i-2n\beta=0 ,\label{BAEgeneqs3}
	\end{align}
\end{subequations}	
where $r_i$ are determined by the Bethe ansatz equations
\begin{equation}\label{BAroots}
	\sum_{j=1,\ j\neq i}^{n} \frac1{r_i-r_j}+\frac{\lambda\,r_i^3+\delta\,r_i^2-\beta\,r_i-2\gamma}{r_i^3}=0 , \qquad i=1,2,\dots,n. 
\end{equation}

\subsection{Solution for arbitrary $n$}
\label{sec:genSol}

In this section we will determine explicit general solutions of our Dirac equation \eqref{eqDirac}. Note that the condition \eqref{BAEgeneqs1} gives the energy relation, so that by substituting $\lambda$ and $\xi_2$ in \eqref{BAEgeneqs1}, after simple manipulations, the energy $\epsilon_n$ can be determined by the following closed-form expression
\begin{equation}\label{nGenEnergy}
	\epsilon_n=\frac{-4 a \alpha ^2 \mathcal{A}^2 b u^2\pm \sqrt{4 \alpha ^2 \mathcal{A}^2 u^2 (a^2-b^2) (\mathcal{B}-\mathcal{A} (2 n+3))^2+(-2 \mathcal{A} n-3 \mathcal{A}+\mathcal{B})^4} }{4 a^2 \alpha ^2 \mathcal{A}^2 u^2+(\mathcal{B}-\mathcal{A} (2 n+3))^2}	,
\end{equation}
in which
\begin{equation}\label{AB}
		\mathcal{A}:=\sqrt{\alpha ^2 S^2 (b-a C)^2} ,\qquad
		\mathcal{B}:=\alpha  (b-a C) \left(2 \alpha  u (b-a C)-3 S\right)  .
\end{equation}
Note that the denominator in \eqref{nGenEnergy} is obviously nonzero (in fact, positive) considering that $a$, $S$ and $\sigma$ are nonzero. Note also that while it seems that the energy expression \eqref{nGenEnergy} depends only on the potential parameter $u$, the other two parameters $v$ and $w$ also play an important role. More precisely, the other two equations \eqref{BAEgeneqs2} and \eqref{BAEgeneqs3} result in severe constraints on the potential parameters as follows
\begin{subequations}\label{vw,nGen} 
	\begin{align}
		v&= \frac{ \alpha  (a C-b) \left(2 \alpha  u (b-a C)-S \Big(n^2+2n+3-2  \sqrt{\frac{1-\epsilon_n ^2}{\alpha ^2}}\sum\limits^{n}_{i=1} r_i \Big)\right)+\mathcal{P}  }{   2 \alpha  S (b-a C) \left(b-\sqrt{1-\epsilon_n ^2} \sqrt{\frac{ (b-a C)^2}{S^2}}+a \epsilon_n \right)  }   ,\\[1ex]  
		w&=  \frac{S \sqrt{\frac{\alpha ^2 (b-a C)^2}{S^2}} \left(\alpha  v (b-a C)+S \Big(2  \sqrt{\frac{1-\epsilon_n ^2}{\alpha ^2}}\sum\limits^{n}_{i=1} r_i^2-(2 n+1)\sum\limits^{n}_{i=1} r_i\Big)\right)+\mathcal{Q}   }{ 2  \sqrt{\alpha ^2(1-\epsilon_n ^2)} (b-a C)^2-2  (a \epsilon_n +b) \sqrt{\alpha ^2 S^2(b-a C)^2} }   \,.
	\end{align}
\end{subequations}	
where
\begin{align*}
    \mathcal{P} &=(n+1) S \sqrt{\frac{\alpha ^2 (b-a cC)^2}{S^2}} (3 S-2 \alpha  u (b-a C))   ,\\[1ex]  
	\mathcal{Q} &= \alpha  (b-a C) \Big(\alpha  (b-a C) \big((2n+1) v+2 u \sum\limits^{n}_{i=1} r_i\big)-3 S \sum\limits^{n}_{i=1} r_i\Big)    \,.
\end{align*}
On the other hand, the wave function associated with the energy \eqref{nGenEnergy}, of \eqref{WavAns} together with \eqref{parameters} and \eqref{ansatZhang}, is given by
\begin{equation}\label{WavAnsnG}
	\begin{aligned}
		\rho_{1,n}(r)&=e^{\Delta(r)}\,\mathcal{R}_{1,n}(r),\\
		\Delta(r)&=\left(\frac32+\frac{\alpha  w (b-a C) (2 \alpha  u (b-a C)-3 S)}{2\sigma  S^2}\right)  \ln r -\frac{\sigma }{r^2} \left(\frac{r v}{w}+\frac{1}{2}\right) -\sqrt{\frac{1-\epsilon_n ^2}{\alpha ^2}}  r .
	\end{aligned}
\end{equation}
In \eqref{vw,nGen}--\eqref{WavAnsnG}, the roots $r_i$ are determined by the following Bethe ansatz equations
\begin{equation}\label{BArootsExpli}
	\sum_{\ j\neq i}^{n} \frac1{r_i-r_j}-\frac{r_i v+w}{r_i^3}\sqrt{\frac{\alpha ^2 (b-a C)^2}{S^2}}+\frac{\alpha  (b-a C) (3 S-2 \alpha  u (b-a C))}{2 r_i\sqrt{\alpha ^2 S^2 (b-a C)^2}}+\frac{3}{2 r_i}-\sqrt{\frac{1-\epsilon_n ^2}{\alpha ^2}}=0 ,
\end{equation}
$i=1,2,...,n$. 
In summary, for a given $n$, the general solutions of the Dirac equation \eqref{eqDirac} are given explicitly by the equations \eqref{nGenEnergy}--\eqref{BArootsExpli}. Below we report explicit solutions for the ground state and the first excited state.

\subsection{Ground-state solution}
\label{sect:Grn0}

For $n=0$ it follows from \eqref{nGenEnergy} that the ground state energy $\epsilon_0$ is given by
\begin{equation}\label{n0Energsimp}
	\epsilon_0= \frac{-4 a \alpha ^2 \mathcal{A}^2 b u^2\pm \sqrt{4 \alpha ^2 \mathcal{A}^2 u^2 (a^2-b^2) (\mathcal{B}-3\mathcal{A})^2+(\mathcal{B}-3 \mathcal{A})^4} }{4 a^2 \alpha ^2 \mathcal{A}^2 u^2+(\mathcal{B}-3\mathcal{A} )^2} \,,
\end{equation}
where $\mathcal{A}$ and $\mathcal{B}$ are given in \eqref{AB}. From \eqref{WavAnsnG} it follows that the associated wave function is
\begin{equation}\label{WavAnsn0}
	\begin{aligned}
		\rho_{1,0}(r)&=c_0\,e^{\Delta(r)},\\
		\Delta(r)&=\left(\frac32+\frac{\alpha  w (b-a C) (2 \alpha  u (b-a C)-3 S)}{2\sigma  S^2}\right)  \ln r -\frac{\sigma }{r^2} \left(\frac{r v}{w}+\frac{1}{2}\right) -\sqrt{\frac{1-\epsilon_0 ^2}{\alpha ^2}}  \,r ,
	\end{aligned}
\end{equation}
where $c_0$ is the normalization constant. Furthermore, from \eqref{vw,nGen}, the potential parameters $v$ and $w$ are given in terms of $u$ as follows
\begin{subequations}\label{vw,n0} 
	\begin{align}
		v&= \frac{  (3 S-2 \alpha  u (b-a C))\left(S \sqrt{\frac{\alpha ^2 (b-a C)^2}{S^2}}+\alpha  (b-a C)\right)} {2 \alpha  S (b-a C) \left(a \epsilon_0 +b-\sqrt{1-\epsilon_0 ^2} \sqrt{\frac{ (b-a C)^2}{S^2}}\right)}  , \\[1ex]  
		w&=  \frac{\alpha   (b-a C) \left(S \sqrt{\frac{\alpha ^2 (b-a C)^2}{S^2}}+\alpha  (b-a C)\right) }{ 2  \sqrt{\alpha ^2(1-\epsilon_0 ^2)} (b-a C)^2-2  (a \epsilon_0 +b) \sqrt{\alpha ^2 S^2(b-a C)^2} }\;  v .
	\end{align}
\end{subequations}	
Note that the Bethe ansatz equations \eqref{BArootsExpli} play no role for the ground state solution since, as established in \eqref{ansatZhang}, $\mathcal{R}_{1,0}(r)\equiv1$.

\subsection{First excited state solution}
\label{sect:Grn01}

For $n=1$, from \eqref{nGenEnergy}, the first excited-state energy, $\epsilon_1$, is  given by
\begin{equation}\label{n1Energsimp}
	\epsilon_1= \frac{-4 a \alpha ^2 \mathcal{A}^2 b u^2\pm \sqrt{4 \alpha ^2 \mathcal{A}^2 u^2 (a^2-b^2) (\mathcal{B}-5\mathcal{A})^2+(\mathcal{B}-5 \mathcal{A})^4} }{4 a^2 \alpha ^2 \mathcal{A}^2 u^2+(\mathcal{B}-5\mathcal{A} )^2} \,,
\end{equation}
where $\mathcal{A}$ and $\mathcal{B}$ are given in \eqref{AB}. The associated wave function, from \eqref{WavAnsnG}, is 
\begin{equation}\label{WavAnsn1}
	\begin{aligned}
		\rho_{1,1}(r)&=c_1(r-r_1)\,e^{\Delta(r)},\\ \Delta(r)&=\left(\frac32+\frac{\alpha  w (b-a C) (2 \alpha  u (b-a C)-3 S)}{2\sigma  S^2}\right)  \ln r -\frac{\sigma }{r^2} \left(\frac{r v}{w}+\frac{1}{2}\right) -\sqrt{\frac{1-\epsilon_1 ^2}{\alpha ^2}}  r ,
	\end{aligned}
\end{equation}
where $c_1$ is the normalization constant. Moreover, from \eqref{vw,nGen}, the potential parameters $v$ and $w$, again both in terms of the Coulombic coefficient $u$, are given by the expressions 
\small
\begin{subequations}\label{vw,n1} 
	\begin{align}
		v&= \frac{ 2 S \sqrt{\frac{\alpha ^2 (b-a C)^2}{S^2}} (3 S-2 \alpha  u (b-a C))-2 \alpha  (b-a C) \left(\alpha  u (b-a C)+S \left(r_1 \sqrt{\frac{1-\epsilon_1 ^2}{\alpha ^2}}-3\right)\right) } {2 \alpha  S (b-a C) \left(a \epsilon_1 +b-\sqrt{1-\epsilon_1 ^2} \sqrt{\frac{ (b-a C)^2}{S^2}}\right)} ,\\[1ex]  
		w&=  \frac{ S \sqrt{\frac{\alpha ^2 (b-a C)^2}{S^2}} \left(\alpha  v (b-a C)+r_1 S \left(2 r_1 \sqrt{\frac{1-\epsilon_1 ^2}{\alpha ^2}}-3\right)\right)+\alpha  (b-a C) (\alpha  (b-a C) (2 r_1 u+3 v)-3 r_1 S) }{ 2  \sqrt{\alpha ^2(1-\epsilon_1 ^2)} (b-a C)^2-2  (a \epsilon_1 +b) \sqrt{\alpha ^2 S^2(b-a C)^2} } \,.
	\end{align}
\end{subequations}	
The unknown parameter $r_1$ appearing in \eqref{WavAnsn1}--\eqref{vw,n1} is determined from \eqref{BArootsExpli} using the following Bethe ansatz equation
\begin{equation}\label{BArootsn1}
	(r_1 v+w)\sqrt{\frac{\alpha ^2 (b-a C)^2}{S^2}}-\frac{r_1^2}{2 }\left(\frac{\alpha  (b-a C) (3 S-2 \alpha  u (b-a C))}{\sqrt{\alpha ^2 S^2 (b-a C)^2}}+3\right)+r_1^3\sqrt{\frac{1-\epsilon_1 ^2}{\alpha ^2}}=0 . 
\end{equation}

\subsection{The special case of Coulomb interaction}
\label{sect:CoulombGrn0}

We will now verify the results obtained by inspecting the special case of the Coulomb interaction, i.e. when $v=w=0$ in \eqref{zr}, and compare the results with Ref.~\cite{Oliviera}. Due to the complicated structure of the potential constraints \eqref{vw,nGen}, it is a rather difficult task to do this for general $n$, but we can check it explicitly for the ground state. From \eqref{vw,n0}, we see that the potential parameters $v$ and $w$ vanish for $(b-a C) \alpha=\frac{3 S}{2u}$. Substituting into $\mathcal{A}$ and $\mathcal{B}$ in \eqref{AB}, we get
	\begin{equation} \label{ABv01}
			\mathcal{A}= \frac{3 S^2}{2\sqrt{u^2}} ,\qquad
			\mathcal{B}= 3 \alpha  S (b-a C)-\frac{9 S^2}{2 u}  .
	\end{equation}
Replacing then $\mathcal{A}$ and $\mathcal{B}$ given by \eqref{ABv01} in \eqref{n0Energsimp} gives that the ground state energy for the Coulomb interaction, denoted by $\epsilon_0^{C}$, is
\begin{equation}\label{n0Energsimpv01u}
		\epsilon_0^{C}(u;a,b,C,S,\alpha)=    \frac{-4 a \alpha ^2 b S^2 u^4\pm u^2 \sqrt{4 \alpha ^2 S^2 \left(a^2-b^2\right) (	\mathcal{F} \left| u\right| +3 S u)^2+(	\mathcal{F}+3 S\, \text{sgn}(u))^4}}{4 a^2 \alpha^2 S^2 u^4+(	\mathcal{F} \left| u\right| +3 S u)^2},	\end{equation}
with $\mathcal{F}:=3 S-2 \alpha  u (b-a C)$. 

Now, to establish a comparison of our ground state energy~\eqref{n0Energsimp} with equation ~(47) of \cite{Oliviera}, we take $a=Z$ and $u=1$, so that\eqref{n0Energsimpv01u} reduces to
	\begin{equation*}\label{n0Energsimpv01}
		\epsilon_0^{C}(1;Z,b,C,S,\alpha)= \frac{-\alpha ^2 b S^2 Z\pm\sqrt{(-\alpha  b+\alpha  C Z+3 S)^2 \left(\alpha ^2 S^2 \left(Z^2-b^2\right)+(-\alpha  b+\alpha  C Z+3 S)^2\right)}}{-6 \alpha  S (b-C Z)+\alpha ^2 (b-C Z)^2+S^2 \left(\alpha ^2 Z^2+9\right)} ,
	\end{equation*}
and then one can easily verify that our ground state energy $\epsilon_0^{C}$ matches equation ~(47) in \cite{Oliviera2019} for $n=2$.

\section{Concluding remarks}

Considering the importance of theoretical studies of curved spacetime in various fields, and based on very recent papers introducing inverse-square and inverse-cubic terms, respectively, as relativistic and quantum corrections to the Coulomb-type gravitational interaction, we consider the Dirac equation in three spatial dimensions with this generalized gravitational potential within a new metric where the curvature effects are embedded in an external field.
The resulting differential equation turns out to be a generalization of the doubly confluent Heun equation, which is an unwieldy equation. 
Using the Bethe-ansatz approach, we address the analytical solution. It should be clear that the Bethe-ansatz technique used here has some limitations and that the problem can only be solved analytically for a certain choice of parameters. In other words, parts of the equations derived in the approach are interpreted as the constraint between the parameters involved. To ensure the validity of the solutions found, the special case of the Coulomb problem was derived and compared with a previous result.

Quite interestingly, the form considered for the scalar potential and the so-called vector (magnetic) potential, is the multipole expansion of these terms. We think this is the first time multipole expansions of the electrostatic and magnetic fields are simultaneously considered for the Dirac equation and an electronic structure.

\appendix

\section{Unsolvability of problem by other approaches}\label{appendix}

From what we have seen throughout the work, it seems that the common exact, quasi-exact or conditionally exact analytical techniques cannot solve our case. Additionally, approaches such as WKB and the quantization rule can be considered to see whether or not acceptable approximate solutions (not to be confused with exact or quasi-exact techniques) are available. Let us now see whether other techniques frequently used  in mathematical physics can solve our problem or not.

\subsection{Known special functions or forms}

The special case of the problem of interest, where the inverse cube term is absent in \eqref{zr}, i.e., $z(r)=\frac{u}{r}+\frac{v}{r^2}$, where $u$ and $v$ are constants, implies a post-Newtonian, or relativistic, correction to the ordinary Coulomb interaction. For such an interaction, the resulting equation has the form
\begin{equation}
R''(r)+\left(\frac{a}{r}+\frac{b}{r^2}+\frac{c}{r^3}+\frac{d}{r^4}+k\right)R(r)=0,
\end{equation}
which is known as the doubly confluent Heun equation and has conditionally (quasi)exact solutions in terms of the doubly confluent Heun functions \cite {El-Jaick, Ronveaux, Ishkhanyan 2018} or by Lie algebra \cite {Turbiner, Turbiner88, Artemio94}.

\subsection{Lie-algebraic approach}

Following the standard idea of quasi-exact solvability \cite{turbiner 2005, turbiner 2015, Turbiner,Turbiner88,Artemio94}, the most general form of a quasi-exactly solvable (QES) differential operator can be represented as a quadratic combination of the $sl(2)$ generators as
\begin{equation}\label{QESLie}
	H_{qes}=\sum_{a,b=0,\pm} C_{ab}\;\mathcal{J}_n^a \mathcal{J}_n^b  +\sum_{a=0,\pm} C_{a}\;\mathcal{J}_n^a +C     \,,
\end{equation}
in which $C_{ab},C_{a},C\in\mathbb{R}$, and the differential operators
\begin{equation}\label{GenerJJJ}
		\mathcal{J}_n^+ =r^2\,\frac{d}{dr}-n\,r ,\qquad
		\mathcal{J}_n^0 = r\,\frac{d}{dr}-\frac n2,\qquad
		\mathcal{J}_n^- = \frac{d}{dr},
\end{equation}
obey the $sl(2)$ commutation relations $[\mathcal{J}_n^+,\mathcal{J}_n^-]=-2\mathcal{J}_n^0$ and $[\mathcal{J}_n^{\pm},\mathcal{J}_n^0]=\mp\mathcal{J}_n^{\pm}$. 
It seems that our case is not expressible in the form of \eqref{QESLie}. To prove this, it suffices to explicitly compute the term $C_{++} \;\mathcal{J}_n^+ \mathcal{J}_n^+$ in \eqref{QESLie}, i.e.
\begin{equation}\label{cplusplus}
	C_{++} \;\mathcal{J}_n^+ \mathcal{J}_n^+=C_{++} \left(r^4\,\frac{d^2}{dr^2}+2(1-n)r^3\,\frac{d}{dr}+n(n-1)r^2 \right),
\end{equation}
which immediately proves the claim: comparing \eqref{cplusplus} with our equation \eqref{eq332}, the coefficient of the second derivative term in \eqref{eq332} indicates that $C_{++}=0$ while the remaining two terms violate this.

\subsection{A class of integral transforms}
The idea of using an integral transform to solve differential  equations is based on solving an equivalent problem in a 
simpler basis and then obtaining the solution of the original problem by an inverse transformation.
The most famous example of this idea in quantum physics is solving the wave equation in momentum space and then recovering the position space solutions by an inverse transformation \cite{Arfken}. 

In this subsection, to convey the main idea, we discuss the Laplace and Fourier transforms, although, generally speaking, a fairly similar argument applies to some other transforms. Before proceeding, it should be noted that the integral transform of special functions is a rather complex field. This is even more complicated in our case, since we are dealing with an equation that appears in a more general form than Heun's. One might think that the Heun functions can be written in a straightforward way in terms of other well-known special functions, such as the Hermite, beta, and hypergeometric functions, however, the story is not that simple even using this idea because of the special form of the associated series. To convey the main idea, let us focus on the basic formulas of the Laplace integral transform, which is usually defined as \cite{Arfken}
\begin{equation}\label{TL}
\mathcal{L}(f(t))=\int_{0}^{\infty} f(t)e^{-st}\,dt =F(s).
\end{equation}
This operation transforms the function $f(t)$ into the function $F(s)$, which in many cases appears in a much simpler form. 
Once the equation for $F(s)$ is solved, we can try to obtain the original function by means of the inverse Laplace transform $\mathcal L^{-1} F(s)=f(t)$. As is well known, from the definition \eqref{TL} it can be easily deduced that the derivative of order $n$ and the multiplication by a power of order $n$ are obtained respectively by
\begin{equation}
\mathcal{L}(f^{(n)}(t))=s^nF(s)-\sum_{k=1}^{\infty}{s^{n-k}}f^{(k-1)}(0^-)
\qquad \text{and} \qquad
\mathcal{L}(t^nf(t))=(-1)^nF^{(n)}(s) .
\end{equation}
Thus, when dealing with a second order differential equation having a polynomial of order $n$ in the effective Hamiltonian in ordinary space, the Laplace transform of the equation appears as a differential equation of order $n$, which at most works smoothly for powers of order two, which of course is not our case.
A rather similar argument applies to the Fourier transform, which is normally used to connect the position and momentum representations of wave functions. To this point we can add the problem of the initial values used in integral transforms, which requires an extensive discussion and is beyond the scope of the present work. To our knowledge, not even integral transforms of Heun functions, which are special cases of our problem, have been adequately addressed in the literature.

\subsection*{Acknowledgments}
The work of MB was supported by the Czech Science Foundation, project 22-18739S.
The research of LMN and SZ was supported by the Q-CAYLE project,
funded by the European Union-Next Generation UE/MICIU/Plan de Recuperacion, Transformacion y Resiliencia/Junta de Castilla y Leon (PRTRC17.11), and also by project PID2023-148409NB-I00, funded by MICIU/AEI/10.13039/501100011033. Financial support of the Department of Education of the Junta de Castilla y Leon and FEDER Funds is also gratefully acknowledged (Reference: CLU-2023-1-05).

%

\end{document}